\documentclass[prd,twocolumn,showpacs,amsmath,amssymb,nofootinbib,superscriptaddress]{revtex4}
\usepackage{graphicx}
\usepackage{epsfig}
\usepackage{bm}
\usepackage{amsfonts}
\usepackage[latin1]{inputenc}
\usepackage{amssymb}
\usepackage{color}
\usepackage{float}
\usepackage{amsmath}                  
\usepackage{dcolumn}
\usepackage{hyperref}

\def\be{\begin{equation}}
\def\ee{\end{equation}}
\def\bea{\begin{eqnarray}}
\def\eea{\end{eqnarray}}
\def\eqi{\begin{equation}}
\def\eqf{\end{equation}}
\def\eqia{\begin{eqnarray}}
\def\eqfa{\end{eqnarray}}

\begin{document}

\title{Linear growth in power law $f(T)$ gravity}

\author{Spyros Basilakos}\email{svasil@academyofathens.gr}
\affiliation{Academy of Athens, Research Center for Astronomy and
Applied Mathematics, Soranou Efesiou 4, 11527, Athens, Greece}


\begin{abstract}
We provide for the first time the growth index of linear matter 
fluctuations of the power law $f(T) \propto (-T)^{b}$ 
gravity model. We find that the asymptotic form of
this particular $f(T)$ model is $\gamma \approx \frac{6}{11-6b}$
which obviously extends that of the $\Lambda$CDM model, 
$\gamma_{\Lambda}\approx 6/11$. Finally, 
we generalize the growth index 
analysis of $f(T)$ gravity in the case where $\gamma$ is allowed 
to vary with redshift.

\end{abstract}

\maketitle


\section{Introduction}
Over the last two decades the 
statistical analysis of cosmological data 
(see Refs.\cite{Hicken2009,Ade2015} and references therein).
supports the idea that the universe is spatially flat and from the overall 
energy density, only $\sim 30\%$ consists 
of matter (luminous and dark). 
Despite the enormous progress made at theoretical 
and observational levels, up to now we know almost nothing 
about the nature of the remaining energy ($\sim 70\%$) and for 
this reason it is given the enigmatic name dark energy (DE). 
The discovery of the physical mechanism 
of dark energy, thought to be driving the 
late accelerated expansion of the universe, is one of 
the main targets 
of theoretical physics and cosmology. In the literature 
one can find 
a plethora of cosmological scenarios that attempt to explain 
the accelerated expansion of the universe. 
In general the cosmological models are mainly classified in two large 
groups. The first category is the so-called scalar field DE models
which adhere to general relativity, proposing however 
the existence of new fields in nature (for review see 
\cite{Ame10}).

Alternatively, models of modified gravity provide an elegant 
mathematical treatment which points 
that the present accelerating epoch appears as a sort
of geometric effect \cite{Ame10}. In this context, 
the corresponding effective 
equation-of-state (EoS) parameter is allowed to take values 
in the phantom regime, namely $w<-1$ 
(for other possible explanations see \cite{Cai10} and \cite{BB14}) 
This situation has been tested 
in WMAP observations, in combination with other 
observational data.
The above feature did not completely disappear 
from the analysis of the Planck data which
indicates that the value of $w$ can 
still be in the phantom region, within 1$\sigma$ uncertainty 
\cite{Ade2015}. For more details concerning the cosmological implications
of modified gravity we refer the reader to the review article of 
Clifton et al. \cite{Clif12}.

Among the large body of nonstandard gravity theories, 
the so-called $f(T)$ gravity 
has been introduced in the literature 
on the basis of the old definition of the so called teleparallel equivalent of
general relativity (TEGR) \cite{ein28,Hayashi79,Maluf:1994ji}.
In the TEGR framework one utilizes the 
corresponding four linearly independent vierbeins and the
curvatureless Weitzenb{\"{o}}ck connection instead of the
torsionless Levi-Civita of the standard General Relativity.
Therefore, the properties of the gravitational 
field are included in the torsion tensor, 
and after performing the appropriate contractions one 
can obtain the torsion scalar $T$ \cite{Hayashi79}. 
Subsequently, inspired by the notations of
$f(R)$ modified gravity, if we allow the Lagrangian of the modified 
Einstein-Hilbert action to be a function of $T$ \cite{Ferraro:2006jd,Ben09,Linder:2010py} then we provide a natural extension of TEGR, 
namely $f(T)$ gravity (for a recent review see \cite{Capft2015}).
The merit of $f(T)$ gravity with respect to $f(R)$ is related to 
the fact that the former produces 
second-order field equations, while the latter 
gives rise to fourth-order equations that may lead to problems, such as 
the well-position and well-formulation of the Cauchy problem \cite{Cap2009}.

But how can we distinguish modified gravity models from those of 
scalar field DE? In order to answer this question we need to test the models
at the perturbation level (for a recent analysis see \cite{Oku2015} and 
references therein). Specifically,
the idea of utilizing the so-called growth index, $\gamma$ 
(first introduced by \cite{Peeb93}),  
of linear matter perturbations as a gravity tool 
is not new and indeed there
is a lot of work in the literature.
There are plenty of studies available in which one can find  
the theoretical form of the
growth index for various cosmological models,
including scalar field DE \cite{Silv94,Wang98,Linjen03,Lue04,Linder2007,Nes08}, 
DGP \cite{Linder2007,Gong10,Wei08,Fu09}, 
Finsler-Randers \cite{Bastav13} and $f(R)$ \cite{Gann09,Tsu09}. 

Despite the fact that the $f(T)$ models have been investigated thoroughly
at the background level 
(see Ref.\cite{Capft2015} and references therein), to the 
best of our knowledge, 
we are unaware of any previous analysis concerning the $f(T)$ growth index. 
In the current article, we wish to study the growth index 
of the power law $f(T) \propto (-T)^{b}$ model \cite{Ferraro:2006jd}.
The layout of the manuscript is as follows: At the beginning of 
Sec. \ref{fTcosmology} we describe the main points of the
$f(T)$ gravity and then we focus our analysis on the 
power law $f(T) \propto (-T)^{b}$ model. In Sec. III we provide
the growth index analysis and the corresponding predictions, 
using two functional forms of the growth index.
Finally, we summarize our conclusions in Sec. IV.

\section{Background expansion in $f(T)$ cosmology}
\label{fTcosmology}
Let us briefly present the basic cosmological properties 
of $f(T)$ gravity. The overall action of $f(T)$ gravity is given by 
\begin{eqnarray}
\label{action11}
 I = \frac{1}{16\pi G_N }\int d^4x e
\left[T+f(T)+L_m+L_r\right],
\end{eqnarray}
where the radiation and matter 
Lagrangians are associated 
with perfect fluids
with pressures $P_r$, $P_m$ and densities $\rho_r$, $\rho_m$  
respectively. Notice, that 
$e = \text{det}(e_{\mu}^A)$ and
${\mathbf{e}_A(x^\mu)}$ are the vierbein fields. 
Within this framework, 
the gravitational field is described by the torsion 
tensor \cite{Hayashi79,Maluf:1994ji} which produces the torsion scalar $T$.
A similar situation holds in the case of the Riemann tensor which 
provides 
the Ricci scalar in standard general relativity.

Considering a spatially flat 
Friedmann-Robertson-Walker (FRW) metric
\begin{equation}
ds^2= dt^2-a^2(t)\,\delta_{ij} dx^i dx^j,
\end{equation}
the vierbien form becomes 
\begin{equation}
\label{weproudlyuse}
e_{\mu}^A={\rm
diag}(1,a,a,a),
\end{equation}
where $a(t)$ is the scale factor of the universe.
Now, if we vary the action (\ref{action11}) with
respect to the vierbeins then we obtain the modified Einstein equations
\begin{eqnarray}\label{eom}
&&e^{-1}\partial_{\mu}(ee_A^{\rho}S_{\rho}{}^{\mu\nu})[1+f_{T}]
 +
e_A^{\rho}S_{\rho}{}^{\mu\nu}\partial_{\mu}({T})f_{TT}\ \ \ \ \  \ \ \ \  \ \
\ \ \nonumber\\
&& \ \ \ \
-[1+f_{T}]e_{A}^{\lambda}T^{\rho}{}_{\mu\lambda}S_{\rho}{}^{\nu\mu}+\frac{1}{4} e_ { A
} ^ {
\nu
}[T+f({T})] \nonumber \\
&&= 4\pi Ge_{A}^{\rho}\overset {\mathbf{em}}T_{\rho}{}^{\nu},
\end{eqnarray}
where $f_{T}=\partial f/\partial T$, $f_{TT}=\partial^{2} f/\partial T^{2}$,
and $\overset{\mathbf{em}}{T}_{\rho}{}^{\nu}$ corresponds to the standard
energy-momentum tensor.

Substituting Eq.(\ref{weproudlyuse}) into the field equations
(\ref{eom}) we derive the Friedmann equations 
\begin{eqnarray}\label{background1}
&&H^2= \frac{8\pi G_N}{3}(\rho_m+\rho_r)
-\frac{f}{6}+\frac{Tf_T}{3}\\\label{background2}
&&\dot{H}=-\frac{4\pi G_N(\rho_m+P_m+\rho_r+P_r)}{1+f_{T}+2Tf_{TT}}.
\end{eqnarray}
In the above set of equations, an overdot denotes
a derivative with respect to time and 
$H\equiv\dot{a}/a$ is the Hubble parameter,
given as a function of torsion $T$ through the 
following equation: 
\begin{eqnarray}
\label{TH2}
T=-6H^2 \;.
\end{eqnarray}
This implies
\begin{eqnarray}
\label{TH3}
E^{2}(a)\equiv\frac{H^2(a)} {H^2_{0}}=\frac{T(a)}{T_{0}},
\end{eqnarray}
where $H_{0}$ is the Hubble constant and $T_0\equiv-6H_{0}^{2}$. 

If we look at the first Friedmann equation
(\ref{background1}) then we realize that it is possible to obtain 
an effective dark energy component. Indeed, it has been shown 
in Ref.\cite{Linder:2010py} that the effective
dark energy density and pressure are given by
\begin{eqnarray}
&&\rho_{DE}\equiv\frac{3}{8\pi
G_N}\left[-\frac{f}{6}+\frac{Tf_T}{3}\right], \label{rhoDDE}\\
\label{pDE}
&&P_{DE}\equiv\frac{1}{16\pi G_N}\left[\frac{f-f_{T} T
+2T^2f_{TT}}{1+f_{T}+2Tf_{TT}}\right],
\end{eqnarray}
where the corresponding effective EoS
parameter is 
\begin{eqnarray}
\label{wfT}
 w=\frac{P_{DE}}{\rho_{DE}}=
-1-\frac{1}{3}\frac{d{\rm ln}T}{d{\rm ln}a}
\frac{f_{T}+2Tf_{TT}}{[(f/T)-2f_{T}]}.
\end{eqnarray}
Combining Eqs.(\ref{TH2}) and (\ref{TH3}) we derive the logarithmic 
derivative of $T$ with respect to $d{\rm ln}a$
\begin{equation}
\label{TDE}
\frac{d{\rm ln}T}{d{\rm ln}a}=2T_{0}E(a)\frac{d{\rm ln}E}{d{\rm ln}a}\;.
\end{equation}

Following standard lines, namely 
$\rho_{m}=\rho_{m0}a^{-3}$ and $\rho_{r}=\rho_{r0}a^{-4}$, 
Eq.(\ref{background1}) is written as
\begin{eqnarray}\label{Mod1Ez}
E^2(a)=\Omega_{m0}a^{-3}+\Omega_{r0}a^{-4}+\Omega_{F0} y(a)
\end{eqnarray}
where
\begin{equation}
\label{LL}
\Omega_{F0}=1-\Omega_{m0}-\Omega_{r0} \;,
\end{equation}
and $\Omega_{i0}=\frac{8\pi G \rho_{i0}}{3H_0^2}$.
Obviously, $f(T)$ gravity affects
the cosmic evolution via the function $y(z)$ (scaled to
unity at the present time), which depends on 
the choice of $f(T)$ as well as on the usual 
cosmological parameters $(\Omega_{m0},\Omega_{r0})$ and it is 
written as 
\begin{equation}
\label{distortparam}
 y(a)=\frac{1}{T_0\Omega_{F0}}\left(f-2Tf_T\right).
\end{equation}

\subsection{Power law model}
In this work we restrict our analysis to the power-law model 
of Bengochea and Ferraro \cite{Ben09}, with
\begin{equation}
\label{Pow}
f(T)=\alpha (-T)^{b},
\end{equation}
where 
\begin{eqnarray}
\alpha=(6H_0^2)^{1-b}\frac{\Omega_{F0}}{2b-1} \;.
\end{eqnarray}
Inserting the above equations into 
Eqs.(\ref{wfT}), and (\ref{distortparam}), we obtain 
\begin{equation}
\label{yLL}
y(a,b)=E^{2b}(a,b) 
\end{equation}
and
\label{TDE1}
\begin{equation}
w=-1-\frac{2b}{3}\frac{d{\rm ln}E}{d{\rm ln}a}=
-1+\frac{2b}{3}(1+z)\frac{d{\rm ln}E}{dz}\;,
\end{equation}
where for the latter equality we have used $a=1/(1+z)$.
In this case the normalized Hubble function (\ref{Mod1Ez}) 
is given by 
\begin{eqnarray}
\label{Mod1Ezz}
E^2(a,b)=\Omega_{m0}a^{-3}+\Omega_{r0}a^{-4}+\Omega_{F0} E^{2b}(a,b) \;.
\end{eqnarray}
Clearly, for $b=0$ the current $f(T)$ model boils down 
to $\Lambda$CDM cosmology\footnote{Notice, that 
for $b=1/2$ it reduces to the Dvali, Gabadadze and Porrati (DGP)
ones \cite{Dvali2000}.}, namely
$T+f(T)=T-2\Lambda$ (where $\Lambda=3\Omega_{F0}H_{0}^{2}$,
$\Omega_{F0}=\Omega_{\Lambda 0}$) and thus we have
\be
E^2(a,0)=\Omega_{m0} a^{-3}+\Omega_{r0} a^{-4}+\Omega_{F0}\equiv E^2_\Lambda(a)
\label{friedlcdm}.
\ee

Notice, that in order to
obtain an accelerating expansion which is consistent with the 
cosmological data one needs $b\ll 1$ \cite{Linder:2010py,Nesseris2013}.
Within this framework, we can now follow the work of 
Nesseris et al. \cite{Nesseris2013}, in which they have shown 
that at the background level all the observationally viable
$f(T)$ parametrizations can be expressed as perturbations deviating to 
$\Lambda$CDM cosmology. In particular,
following the notations of \cite{Nesseris2013}  
for the power law $f(T)$ model 
we perform a Taylor expansion of $E^2(a,b)$ around $b=0$ 
$$
E^2(a,b)=E^2(a,0)+\left.\frac{dE^2(a,b)}{db}\right|_{b=0} b+... 
$$
or
\be
E^2(a,b)=E^2_{\Lambda}(a)+\Omega_{F0}\left.\frac{dy(a,b)}{db}\right|_{b=0}
b+...\;,
\label{ModapE}
\ee
where for the latter equality we have used Eq.(\ref{distortparam}).
Now based on Eq.(\ref{yLL}) we obtain 
\be
\frac{dy(a,b)}{db}=2E(a,b)^{2 b} \left\{\frac{b }{E(a,b)} \frac{dE(a,b)}{db}+ \ln
\left[E(a,b)\right]\right\},
\ee
and evaluating the above equation for $b=0$ we find
\be
\label{yya}
\frac{dy(a,b)}{db}|_{b=0}=2\ln\left[E(a,0)\right]=\ln
\left[E^2_\Lambda(a)\right].
\ee
Therefore, inserting Eq.(\ref{yya}) into 
Eq.(\ref{ModapE}) we provide
the approximate normalized Hubble parameter for the 
current $f(T)$ model (see \cite{Nesseris2013})
\be
\label{approxM1}
E^2(a,b)\simeq E^2_\Lambda(a)+\Omega_{F0}\ln\left[E^2_\Lambda(a)\right]b \;.
\ee

Implementing an overall likelihood analysis involving the
latest cosmological data 
(SNIa \cite{Suzuki:2011hu}, BAO \cite{Blake:2011en,Perc10} 
and Planck CMB shift parameter \cite{Shaef2014}) 
and the appropriate Akaike information criterion
\cite{Akaike1974} we can place constraints 
on the cosmological parameters $(\Omega_{m0},b)$.
Specifically, we find that the likelihood function peaks at
$\Omega_{m0}=0.286\pm 0.012$, $b=-0.081\pm 0.117$ with 
$\chi^{2}_{\rm min}(\Omega_{m0},b)\simeq 563.6$ (AIC=567.6), resulting in 
a reduced value of $\sim 0.96$.\footnote{The total $\chi^{2}$ function  
is given by $\chi^{2}=\chi^{2}_{\rm SNIa}+\chi^{2}_{\rm BAO}+\chi^{2}_{\rm CMB}$.
For Gaussian errors, the 
Akaike information criterion (AIC) \cite{Akaike1974} is given by
${\rm AIC}=\chi^2_{t,min}+2k$, where 
$k$ provides the number of free parameters.}
In order to visualize the solution space 
in Fig.1 we plot the 1$\sigma$, 2$\sigma$ and $3\sigma$
confidence contours in the $(\Omega_{m0},b)$ plane.
At this point we need to mention that the uncertainty
of the $b$ parameter 
is quite large (see also \cite{Nesseris2013}), as indicated in 
the relevant contour figure.
Our statistical results are in agreement, within $1\sigma$ errors,
with those of Nesseris et al. \cite{Nesseris2013}, who used 
a combined analysis of  
SNIa \cite{Suzuki:2011hu}, BAO \cite{Blake:2011en,Perc10} 
and WMAP9 CMB shift parameters \cite{Hinshaw:2012fq} and they 
found $(\Omega_{m0},b)=(0.274\pm 0.008,-0.017\pm 0.083)$.

For the concordance $\Lambda$ cosmology ($b=0$) we find 
$\Omega_{m0}=0.289\pm 0.012$, 
$\chi^{2}_{\rm min}(\Omega_{m0})\simeq 564.6$ (AIC=566.6).
Since the difference $|\Delta {\rm AIC}|$=$|{\rm
AIC}_{\Lambda}-{\rm AIC}_{f(T)}|< 2$ points to the fact that the power law 
$f(T)$ and $\Lambda$CDM models respectively 
fit the cosmological data equally well.

\begin{figure}[t]
\includegraphics[width=0.5\textwidth]{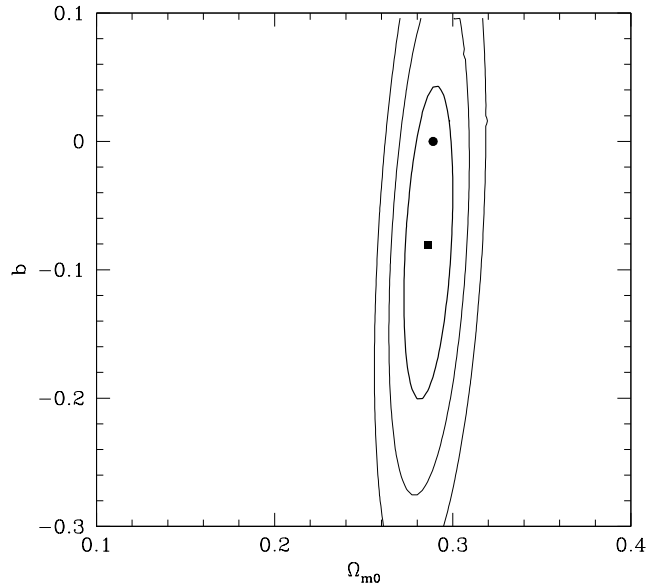}
\caption{The overall (SNIa/BAO/CMB$_{shift}$) likelihood
contours for $\Delta \chi^2=\chi^{2}-\chi^{2}_{\rm min}$ equal to
1$\sigma$ (2.32), 2$\sigma$ (6.18) and 
3$\sigma$ (11.83) confidence levels, in the $(\Omega_{m0},b)$ plane. 
The solid square corresponds to the best-fit $f(T) \propto (-T)^{b}$ modified 
gravity model, namely 
$(\Omega_{m0},b)=(0.286,-0.08)$. The solid point shows the best-fit 
solution for the concordance $\Lambda$CDM model.}
\end{figure}


\section{Linear growth in $f(T)$ cosmology}
In this section we present the linear matter fluctuations of $f(T)$ gravity
in the matter dominated era (for details see Ref.\cite{Li2011}).
Therefore, for the rest 
of the paper we neglect the radiation term from the cosmological 
expressions appearing in section II.
Based on standard treatment, 
the differential equation that describes the evolution 
of matter perturbations at the sub-horizon scales takes on the form
\begin{equation}
\label{eq:111}
\ddot{\delta}_{m}+2\nu H\dot{\delta}_{m}-4\pi G\mu \rho_{m} \delta_{m}=0 \;.
\end{equation}
In the framework of modified gravity models the quantity $\mu=G_{\rm eff}/G_{N}$ 
depends on the scale factor, while for those dark energy models which 
are inside general relativity $G_{\rm eff}$ reduces to Newton's constant
as it should and thus $\mu=1$. 
We refer the reader to Refs.
\cite{Lue04,Linder2007,Gann09,Stab06,Uzan07,Tsu08,Steigerwald:2014ava} for full details of the calculation.
One can show that $\delta_{m} \propto D(t)$ where $D(t)$
is the linear growth factor scaled to unity at the present epoch.
Obviously, any modification to the gravity theory 
and to the Friedmann equation
is reflected in the quantities $\nu$ and $\mu\equiv G_{\rm eff}/G_{N}$.
As an example, in the framework of scalar field dark energy models
which adhere to general relativity one has $\nu=\mu=1$. 
Moreover, for the concordance $\Lambda$ cosmology, one
can solve  (\ref{eq:111}) analytically in order to obtain
the growth factor \cite{Peeb93}
\begin{equation}
\label{eq24}
D_{\Lambda}(a)=\frac{5\Omega_{m0}
  E_{\Lambda}(a)}{2}\int^{a}_{0}   
\frac{du}{uE^{3}_{\Lambda}(u)},
\end{equation}
where
\begin{equation}
\label{ELL}
E_{\Lambda}(a)=\left( \Omega_{m0}a^{-3}+\Omega_{\Lambda0}\right)^{1/2}
\end{equation}
in the matter dominated era and $\Omega_{\Lambda0}=1-\Omega_{m0}$.

On the other hand 
for nonstandard gravity models we have $\nu=1$ and $\mu \ne 1$
and for the $f(T)$ gravity the quantity $\mu$ 
takes the following form \cite{Zheng:2010am,Chen001}:
\begin{eqnarray}
\label{Geff}
\mu=\frac{1}{1+f_{T}}.
\end{eqnarray}
Inserting Eq.(\ref{Pow}) into Eq.(\ref{Geff}) we obtain
\begin{equation}
\mu(a)=\frac{1}{1+ \frac{b\Omega_{F0}} {(1-2b)E^{2(1-b)}}}
\end{equation}
or
\be
\label{Geff1}
\mu(a)\simeq 1-\frac{\Omega_{F0}}{E^{2}_{\Lambda}(a)}\;b+...
\ee
where, as in section II, for the latter expression we have 
utilized a Taylor expansion around $b=0$.

In order to simplify the numerical calculations we provide the 
growth rate of clustering introduced 
by \cite{Peeb93}
\begin{equation}
\label{fzz221}
f(a)=\frac{d\ln \delta_{m}}{d\ln a}\simeq \Omega^{\gamma}_{m}(a),
\end{equation}
based on which we can write the growth factor
\be
\label{eq244}
D(a)={\rm exp} \left[\int_{1}^{a}
\frac{\Omega_{m}(x)^{\gamma(x)}}{x} dx \right]\;,
\ee
with 
\begin{equation}
\label{ddomm}
\Omega_{m}(a)=\frac{\Omega_{m0}a^{-3}}{E^{2}(a)} 
\end{equation}
and from which we define
\begin{equation}
\label{ddomm1}
\frac{d\Omega_{m}}{da}=-3\frac{\Omega_{m}(a)}{a}\left( 1+\frac{2}{3}
\frac{d{\rm ln}E}{d{\rm ln}a} \right) \;.
\end{equation}

The parameter $\gamma$ is the so-called growth index which 
can be used to distinguish between general
relativity and modified gravity on cosmological scales (see Introduction). 
In this context, utilizing the first equality of (\ref{fzz221}) one can
write Eq.(\ref{eq:111}) as follows:
\be \label{fzz444} 
a\frac{df}{da}+
\left(2\nu+\frac{d{\ln}E}{d{\rm ln}a}\right)f+f^{2}
=\frac{3\mu \Omega_{m}}{2}\;. 
\ee 
Now differentiating Eq.(\ref{Mod1Ezz}) and utilizing
Eq.(\ref{ddomm}) we find that
\be
\frac{d{\ln}E}{d{\rm ln}a}=
-\frac{3}{2}\frac{\Omega_{m}(a)}{[1-bE^{2(b-1)}\Omega_{F0}]} \;.
\ee
For $b\ll 1$ the latter equation is well approximated by 
\be
\label{Taylor2}
\frac{d{\ln}E}{d{\rm ln}a}\simeq 
-\frac{3}{2}\Omega_{m}(a)\left[ 1+\frac{\Omega_{F0}b}{E^{2}_{\Lambda}(a)}+...\right]
\ee

Regarding the form of the growth index we consider the following two 
situations. 

\subsection{Constant growth index}
The simplest choice is to use the asymptotic value
of the growth index, namely $\gamma_{\infty}$.
Recently, Steigerwald et al. \cite{Steigerwald:2014ava} 
proposed a general mathematical treatment 
which provides $\gamma_{\infty}$ analytically
(see Eq.(8) in \cite{Steigerwald:2014ava} and the discussion 
in \cite{Basola}) 
for a large family of DE models. 
Based on the work of Steigerwald et al. \cite{Steigerwald:2014ava} 
the asymptotic 
value of the growth index is given analytically by
\be
\label{g000}
\gamma_{\infty}=\frac{3(M_{0}+M_{1})-2(H_{1}+N_{1})}{2+2X_{1}+3M_{0}}
\ee
where the relevant quantities are 
\be \label{Coef1}
M_{0}=\left. \mu \right|_{\omega=0}\,,
\ \
M_{1}=\left.\frac{d \mu}{d\omega}\right|_{\omega=0}
\ee
and
\be \label{Coef2}
N_{1}=\left.\frac{d \nu}{d\omega}\right|_{\omega=0}\,,\ \
H_{1}=-\frac{X_{1}}{2}=\left.\frac{d \left(d{\rm ln}E/d{\rm ln}a\right)}{d\omega}\right|_{\omega=0} \,.
\ee
We would like to point out that 
Steigerwald et al. \cite{Steigerwald:2014ava} derived
the basic cosmological functions in terms of the variable
$\omega={\rm ln}\Omega_{\rm m}(a)$, 
which implies that at $z\gg 1$ we have 
$\Omega_{m}(a)\to 1$ [or $\omega \to 0$]\footnote{For $\Lambda$ cosmology ($b=0$) 
Eq.(\ref{Taylor2}) becomes 
$\frac{d{\rm ln}E_{\Lambda}}{d{\rm ln}a}=-\frac{3\Omega^{(\Lambda)}_{m}(a)}{2}$, 
where $\Omega^{(\Lambda)}_{m}(a)=\frac{\Omega_{m0}a^{-3}}{E^{2}_{\Lambda}(a)}$.
Of course at large redshifts $z\gg 1$ we have 
$\Omega^{(\Lambda)}_{m}(a) \to 1$ and thus 
$\frac{d{\rm ln}E_{\Lambda}}{d{\rm ln}a}\to -\frac{3}{2}$.}.
For the $f(T)$ gravity the coefficient 
$N_{1}$ is strictly equal to zero since $\nu=1$.
Then, based on Eqs.(\ref{approxM1}), (\ref{Geff1}), (\ref{ddomm}), 
(\ref{ddomm1}) and
(\ref{Taylor2}), we find after 
some algebra (for more details see the Appendix)
$$
\{ M_{0},M_{1},H_{1},X_{1}\}\simeq \{ 1,b,-\frac{3(1-b)}{2},3(1-b)\}
$$
and thus we calculate for 
the first time (to the best of our knowledge) the
asymptotic value of the growth index
\be
\label{g002}
\gamma_{\infty} \simeq \frac{6}{11-b} \approx 
\frac{6}{11}\left( 1+\frac{6}{11}b\right) \;.
\ee
Obviously, for $b=0$ 
we recover the $\Lambda$CDM value $6/11$ as we should.
On the other hand, utilizing the aforementioned 
best-fit value $b=-0.081$ and the corresponding $1\sigma$ $b-$uncertainty 
$\sigma_{b}=0.117$, we find
that $\gamma_{\infty}$ lies in the interval  
$[0.492,0.556]$ (see upper panel of Fig.2).
In the lower panel of Fig.2 we show the relative deviation  
of the $f(T)$ growth index with respect to
$\gamma_{\Lambda} \approx 6/11$.
The relative difference can reach $\sim -9\%$ when we approach
the aforesaid theoretical lower $1\sigma$ bound of $b \simeq -0.2$. 
For the best fit value $b=-0.081$ we have $\gamma=0.5223$ that 
gives a $\sim -4\%$ difference from $6/11$.
We also see that for positive values of $b$ the asymptotic value 
of the growth index becomes greater 
than $6/11$, while the opposite holds for negative values of $b$.

\begin{figure}
\mbox{\epsfxsize=8.2cm \epsffile{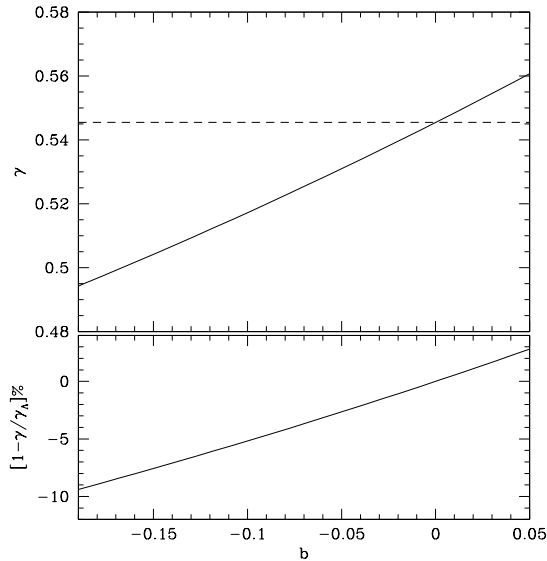}} \caption{{\it Upper panel:}
We show the asymptotic value of the growth
index as a function of $b$ (solid line).
The dashed curve corresponds to $\gamma_{\Lambda}\approx 6/11$.
{\it Lower panel:} We plot the relative difference 
$[1-\gamma/\gamma_{\Lambda}]\%$ versus $b$.}
\end{figure}

\subsection{Varying growth index}
The second possibility is to consider that $\gamma$ evolves with redshift.
Therefore, in this scenario we need to generalize the 
original Polarski and Gannouji \cite{Pol} method for the $f(T)$ gravity. 
Specifically, upon substituting $f(a)=\Omega_{m}(a)^{\gamma(a)}$ 
into Eq.(\ref{fzz444}) and using Eq.(\ref{ddomm1})
we are led to
{\small{
\begin{equation}
\label{Poll}
a{\rm ln}(\Omega_{m})\frac{d\gamma}{da}+\Omega_{m}^{\gamma}
-3\left(\gamma-\frac{1}{2}\right)\left(1+\frac{2}{3}
\frac{d{\ln}E}{d{\rm ln}a}\right)+\frac{1}{2}=\frac{3}{2}
\mu\Omega_ { m } ^ { 1-\gamma},
\end{equation}}}
Writing the above equation at the present time ($a=1$) we simply have 
\begin{eqnarray}
\label{Poll1}
&&-\gamma^{\prime}(1){\rm
ln}(\Omega_{m0})+\Omega_{m0}^{\gamma(1)}-3\left[\gamma(1)-\frac{1}{2}\right]
\left(1+\frac{2}{3}\frac{d{\ln}E}{d{\rm ln}a}\right)_{a=1}\nonumber\\
&&
+\frac{1}{2}=\frac { 3
} { 2 } \mu_ { 0 } \Omega_{m0}^{1-\gamma(1)}, \ \ \
\end{eqnarray}
where a prime denotes a derivative with respect to the scale factor
and
$$\mu_{0}=\mu(1)\simeq 1-\Omega_{F0}b,$$

$$\left.\frac{d{\ln}E}{d{\rm ln}a}\right|_{a=1}\simeq -\frac{3}{2}\Omega_{m0}(1+\Omega_{F0}b)$$ 
For the latter two expressions we have used
Eqs.(\ref{Geff1}) and (\ref{Taylor2}).

In this work we consider the most popular $\gamma(a)$ parametrization  
that has appeared in the literature (see \cite{Pol,Bel12,DP11,Ishak09,Bass}), 
which is a Taylor expansion around $a(z)=1$
\be
\label{aPoll1}
\gamma(a)=\gamma_{0}+\gamma_{1}(1-a)\;,
\ee
with the asymptotic value becoming 
$\gamma_{\infty}\simeq \gamma_{0}+\gamma_{1}$ where we have set 
$\gamma_{0}=\gamma(1)$.

Utilizing Eqs.(\ref{Poll1}), and (\ref{aPoll1}), 
and the above 
notations we can easily obtain
$\gamma_{1}$ in terms of $\gamma_{0}$:
\begin{equation}
\label{Poll2}
\gamma_{1}=\frac{\Omega_{m0}^{\gamma_{0}}-3(\gamma_{0}-\frac{1}{2})
[1-\Omega_{m0}(1+\Omega_{F0}b)]-\frac{3}{2}\mu_{0}\Omega_{m0}^{1-\gamma_{0}}+\frac{1}{2}  }
{\ln  \Omega_{m0}}\;.
\end{equation}
As expected, for the $\Lambda$ cosmology ($b=0$)
the above formula reduces to its standard expression
\cite{Pol,Bel12,DP11,Ishak09,Bass}.
Lastly, inserting 
$\gamma_{0}=\gamma_{\infty}-\gamma_{1}$ into
Eq.(\ref{Poll2}) and utilizing 
$\gamma_{\infty}\approx \frac{6}{11-6b}$ we can derive the constants
$\gamma_{0,1}$ as a function $(\Omega_{m0},b)$. 
For example, if we use the fitting values 
$(\Omega_{m0},b)=(0.286,-0.081)$ then we estimate 
$(\gamma_{0},\gamma_{1}) \simeq (0.541,-0.019)$, while for the concordance
$\Lambda$ cosmological model with $(\Omega_{m0},b)=(0.289,0)$
we find 
$(\gamma_{0},\gamma_{1})\simeq (0.557,-0.011)$.

In order to check the variants of the $f(T) \propto (-T)^{b}$ 
model from the $\Lambda$CDM case at the perturbation level
we present in Fig.3 a comparison of the 
evolution of the growth index $\gamma(z)$ (upper panel) and 
the evolution of the $\mu(z)\equiv G_{\rm eff}(z)/G_{N}$ (lower panel). 
The solid and the dashed curves correspond to $f(T)$ and 
$\Lambda$CDM models respectively.
Also, the thin-line error bars correspond to 
$1\sigma$ $b$-uncertainties which affect 
the growth index and $\mu$ via Eqs.(\ref{Geff1})
and (\ref{Poll2}). As expected, at large redshifts $f(T)$ tends to general 
relativity, namely $\mu \to 1$, while
as we approach the present epoch $\mu$ starts to deviate from unity. 
Of course, due to large $1\sigma$ $b$-uncertainties we cannot 
exclude the value $b=0$ which corresponds to the 
concordance $\Lambda$ cosmology.
Therefore, in order to test possible 
departures from general relativity 
we need to place tight constraints 
on the $b$ parameter and thus on $\gamma$.
Hopefully, using the next generation of surveys (like {\em Euclid} 
see discussion in \cite{Sapone13}) we expect to be able to constrain 
the $b$ parameter.

\begin{figure}
\mbox{\epsfxsize=8.2cm \epsffile{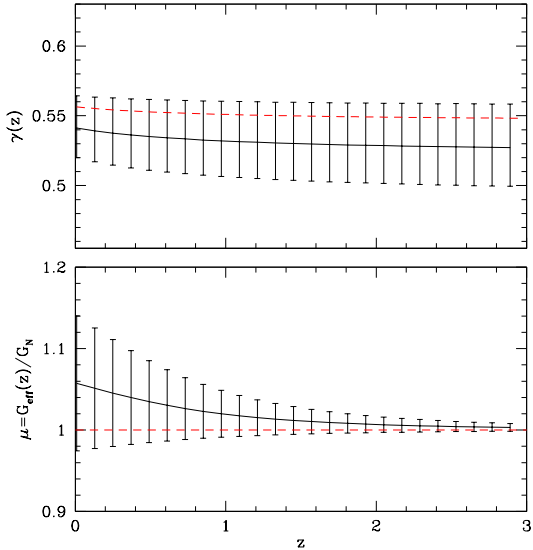}} \caption{In the upper panel
we show the growth index as a 
function of redshift for the $f(T)\propto (-T)^{b}$ 
gravity model (solid line).
In the lower panel we plot the evolution of the $\mu(z)\equiv G_{\rm eff}/G_{N}$
[see Eq.(\ref{Geff1})].
Notice, that the thin-line error bars correspond to 
$1\sigma$ $b$-uncertainties which affect 
the growth index and $\mu$ via Eqs.(\ref{Poll2})
and (\ref{Geff1}).
For comparison, the dashed line
corresponds to the traditional $\Lambda$CDM model.}
\end{figure}



\section{Conclusions}
\label{conclusions}
We studied the power-law $f(T) \propto (-T)^{b}$ model at the linear 
perturbation level. Applying the technique 
of Steigerward et al. \cite{Steigerwald:2014ava} 
in the framework of the current $f(T)$ model we 
derive (for the first time) the asymptotic value of the 
growth index of matter perturbations, namely $\gamma\approx \frac{6}{11-6b}$.
Evidently, for $b=0$ the latter formula reduces to that of the 
usual $\Lambda$CDM model, $\gamma_{\Lambda}\approx \frac{6}{11}$.
It is interesting to mention that Nesseris et al. \cite{Nesseris2013}
proved that 
the power-law $f(T)$ model can be seen as a perturbation around 
$\Lambda$CDM at the expansion level. Here we extended the latter work,
by writing the asymptotic value of the $f(T)$ growth index 
as a perturbation around that of $\Lambda$CDM, namely 
$\gamma \approx \frac{6}{11}\left( 1+\frac{6}{11}b\right)$
Finally, we generalized the analysis in the regime where
the growth index is allowed to vary with redshift and we found that 
an accurate determination of $b$ is needed in order to test the 
range of validity of the $f(T) \propto (-T)^{b}$ modified gravity
model.

\section*{Acknowledgements}
I would like to acknowledge 
support by the Research Center for Astronomy of the Academy
of Athens in the context of the program ``{\it Tracing the Cosmic
Acceleration}''. \\

\appendix
\section{Basic coefficients} 
Here we provide some 
calculations concerning the coefficients
$M_{0}$, $M_{1}$, $H_{1}$ and $X_{1}$ which 
appear in Eq(\ref{g000}). As we have already discussed in section 
IIA, these quantities are given in terms of 
the variable $\omega={\ln}\Omega_{m}$ 
which means that as long as $a\to 0$ ($z\gg 1$) we have $\Omega_{m}\to 1$ 
(or $\omega \to 0$) and thus $E^{2}(a) \gg 1$.
Therefore, from Eq.(\ref{Geff1}) we simply find 
$$
M_{0}=\left. \mu \right|_{\omega=0} \simeq 1 \;.
$$
Now, $M_{1}$ is defined
as
$$
M_{1}=\left.\frac{d \mu}{d\omega}\right|_{\omega=0}=
\left.\Omega_{m}\frac{d \mu}{d\Omega_{m}}\right|_{\Omega_{m}=1} .
$$
Using Eqs.(\ref{ELL}), (\ref{Geff1}), (\ref{ddomm}), and (\ref{ddomm1}) we 
obtain after some calculations 
\begin{equation*} 
\Omega_{m}(a)\frac{d \mu}{d\Omega_{m}}
\simeq \Omega_{m}(a)
\frac{b\Omega_{F0}}{E^{2}_{\Lambda}(a)\Omega_{\Lambda}(a)}
=\Omega_{m}(a)\frac{b\Omega_{F0}}{\Omega_{\Lambda0}}\;.
\end{equation*}
Notice, that for the latter equality we use the 
well-known formula $E^{2}_{\Lambda}(a) \Omega_{\Lambda}(a)=\Omega_{\Lambda0}$.
Under of these conditions $M_{1}$ becomes
$$
M_{1} \simeq \frac{b\Omega_{F0}}{\Omega_{\Lambda0}} \simeq b,
$$
where we have set $\Omega_{F0}=\Omega_{\Lambda0}$ 
[see the corresponding discussion before Eq.(\ref{friedlcdm})].

Finally, the coefficient $H_{1}$ (or $X_{1}$) is given by
$$
H_{1}=-\frac{X_{1}}{2}=\left.\frac{d \left(d{\rm ln}E/d{\rm ln}a\right)}{d\omega}\right|_{\omega=0} =
\left.\Omega_{m}\frac{d \left(d{\rm ln}E/d{\rm ln}a\right)}{d\Omega_{m}}\right|_{\Omega_{m}=1} .
$$
Again,
utilizing 
Eqs.(\ref{ELL}), (\ref{Geff1}), (\ref{ddomm}), (\ref{ddomm1}) 
and (\ref{Taylor2})
we find 
\begin{equation*} 
\Omega_{m}\frac{d \left(d{\rm ln}E/d{\rm ln}a\right)}{d\Omega_{m}}\simeq 
-\frac{3\Omega_{m}}{2}\left[1+\frac{b\Omega_{F0}}{E^{2}_{\Lambda}(a)}
+\frac{2b\Omega_{F0}}{3E^{2}_{\Lambda}(a)\Omega_{\Lambda}(a)}
\frac{d{\rm ln}E_{\Lambda}}{d{\rm ln}a}.
\right]
\end{equation*} 
Therefore, in the context of the aforementioned limitations ($\Omega_{m} \to 1$)
$H_{1}$ (and thus $X_{1}$) takes the form  
$$
H_{1}=-\frac{X_{1}}{2}\simeq -\frac{3}{2}(1-b).
$$

\end{document}